\documentclass[aps,prl,reprint,groupedaddress]{revtex4}
\usepackage{graphics,graphicx}
\usepackage{epsfig}
\begin{document}

%\preprint{}

%\title{ Little-Higgs-like-Dark Matter (

\title{A SM Singlet Scalar as Dark Matter}

 \author{Tonatiuh Matos\footnote{Part of the Instituto Avanzado de Cosmolog\'ia (IAC)
  collaboration http://www.iac.edu.mx/}}
 \email{tmatos@fis.cinvestav.mx}
 \author{Ricardo Lopez-Fernandez}
 \email{rlopez@fis.cinvestav.mx}
 \affiliation{Departamento de F\'isica, Centro de Investigaci\'on y de Estudios
Avanzados del IPN, A.P. 14-740, 07000 M\'exico D.F.,
             M\'exico.}
%\thanks{}
%\affiliation{aqui}


%\date{\today}

\begin{abstract}
 In this work we investigate the possibility that a simple extension of the
 Standard Model (SM) can be the dark matter of the universe. We postulate
 the  existence of  a scalar field singlet like the Higgs as an extra term
 in the SM Lagrangian. We find that from the astrophysical point of view 
 a very small mass and self-interaction is more convenient to agree with 
 observations and from particle detectors observations we do not see any essential 
 constrain to this settings. Thus, we conclude that a scalar field singlet
 with a small mass and self-interaction is a good candidate to be the
 nature of the dark matter.
\end{abstract}

\pacs{}
%\keywords{}

\maketitle

%\section{\label{sec:level1}INTRODUCTION}
%First-level heading:\protect\\ The line break was forced \lowercase{via}\textbackslash\textbackslash}
%\subsection{\label{sec:level2}Second-level heading: Formatting}
%\subsubsection{}

\section{Introduction}

In the last time the most accepted candidates to be the dark matter of the universe
have been ruled out by observations. The supersymmetric candidate have been
extremely constrained by the LUX \cite{lux} detector and recently the axion field  
seems to be in tension with the BICEP2 results \cite{marsh}.
 The accepted scientific paradigm for understanding the evolution 
of the universe is based on the theory of general relativity and the standard model 
of particles. Nevertheless, the discovery that the universe is filled out in more 
than 96\% with two unknown kind of matters is putting these two theories in alert. 
Nowadays there is no doubt that we need to modify or extend one or both of these 
theories in order to explain the existence of the dark matter (DM) and the dark 
energy. Of course, the first idea is to modify one of them. There are hundreds  
of papers studying the modifications of these theories in order to give one 
explanation of these two kind of matters. The modification of the standard model 
give rise to the existence of new particles, the most accepted candidates are 
neutralinos, gravitinos, higgsinos, etc. From the modification of the theory 
of general relativity we also have several very interesting proposals. There 
is a third way to explain the existence of the dark sector of the universe. 
We can also postulate that there exist extra interactions in the universe, 
besides the strong, weak and electromagnetic fundamental interactions in the 
standard model, all of them of spin one, and the gravitational interaction of 
spin two. In this work we propose to explore this way supposing that the DM is 
the consequence of the existence of a new interaction in the universe. If this 
is the case, this interaction needs to be a boson. We start with the most simple 
case of an interaction of spin zero. It most be of long range only, but it must 
let the rest of the interactions intact at small scales. This implies that the 
mass of this zero boson is very small. And it must interact very tiny or not at 
all with the rest of the matter. If this is the case, we have to add to the 
Lagrangian of the standard model plus general relativity the contribution of 
this interaction as
\begin{equation}
{\cal L}={\cal L}_{GR}+{\cal L}_{SM}+\frac{1}{2}(\partial_\mu \Phi)^2+V+\alpha{\cal
L}_{Int}
\label{eq:Lagran}
\end{equation}
where ${\cal L}_{GR}$ represents the Lagrangian of General Relativity, ${\cal
L}_{SM}$ the one of the standard model and ${\cal L}_{Int}$ the Lagrangian for the
interaction between matter and the scalar field $\Phi$. In order that the scalar
field mimics the DM, it is convenient that the scalar field potential $V$ remains
convex. The simplest potential with this features is the ``Higgs" potential
$V=-1/2m^2\Phi^2+\lambda/4\Phi^4+\cdots$. For simplicity we can start with a real
scalar field, but the results presented here are also valid for a complex one.
 
 Observe that the scalar field in Lagrangian (\ref{eq:Lagran}) could be interpreted
as an extra particle in the standard model as well. In any case, both
interpretations of this particle are a very simple modification of the standard
model of particles.
 
 The main goal of this work is to constrain the values of the three constants of
this model $m$, $\lambda$ and $\alpha$ using actual observations in particle
physics and in astrophysics. 

\section{Astrophysical Contrains}

From the astrophysical point of view this scalar field can be constrained using the
resent observations of the Planck satellite, the supernova observations, the
rotation curves of galaxies, etc. 

We start analyzing the behavior of the scalar field at cosmological scales. The idea
is the following, at very high temperature the scalar field interacts with the rest
of matter. This interaction can be mimicked supposing that the scalar field lies in
a thermal bath. The scalar field potential can be approximated in a thermal bath at
one loop as
\begin{equation}
V=\frac{\lambda}{4}\left(\Phi^2-\frac{m^2}{\lambda}\right)^2+\frac{1}{8}T^2\Phi^2-\frac{\pi^2}{90}T^4
\label{eq:V}
\end{equation}
where $T$ is the temperature of the thermal bath. This ``Higgs" potential is very
well know, it is convex when $T$ is bigger than the symmetry breaking temperature
$T_{SB}=2m/\sqrt{\lambda}$, and it has the mexican hat shape when $T<T_{SB}$. As the
universe expands, it riches the decoppling temperature of the scalar field with the
rest of matter. As the temperature keeps going down, the scalar field riches the
temperature of symmetry breaking at $T_{SB}$, the $Z_2$ symmetry breaks down and the
scalar field goes into the second minimum of the potential $V$ at
$\Phi_{min}=\sqrt{m^2/\lambda-T^2/4}=1/2\sqrt{T_{SB}^2-T^2}$. The scalar field
starts to oscillate around the second minimum with a frequency $\sim m$. Here it is
convenient to analyze the dynamic of the scalar field moving the origin of potential
to the new minimum using the transformation $\Phi\rightarrow\Phi-\Phi_{min}$, such
that we can neglect the higher order terms, thus the potential $V$ in this new
coordinates looks like a $\Phi^2$ scalar field potential. The mass $M$ of the scalar
field at this point is $M=\sqrt{2}m$. 

Thus, potential $V$ guaranties that the scalar field stabilizes oscillating around
its second minimum. In the oscillation state the scalar field behaves as dust and it
mimics very well the Cold Dark Matter (CDM). As in the case of the CDM, small
fluctuations of the scalar field collapse, the size of the collapse depends of the
mass of the scalar field. For a big mass $M$, it forms small stars and for small
mass $M$ it forms big stars, in other words, gravity confines the scalar field in
specific regions where the scalar field starts to collapse. 
Because of the tiny coupling constant the scalar field behaves as an ideal gas,
which critical temperature of condensation is 
\begin{equation}
T_c=\frac{2\pi}{M^{5/3}}\left(\frac{\rho}{2.612}\right)^{\frac{2}{3}}
\label{eq:Tc}
\end{equation}
where $\rho$ is the density of the scalar field in the specific volum $v$ where it
is confined. For a tiny mass of order of eV the critical mass of condensation is of
the order of TeV. Thus we expect that the scalar field forms Bose-Einstein
Condensates (BEC) very early in the universe. Furthermore, the critical mass of
collapse is
\begin{equation}
M_c=0.06\sqrt{\lambda}\frac{m_{pl}^3}{M^{2}}
\label{eq:Mc}
\end{equation}
where $m_{pl}$ is the Planck mass. This critical mass $M_c$ can be interpreted as a
maximal mass of collapse, that means that the scalar field will form self
gravitating objects with mass $M_c$ and lower. Thus, for example, axions have a mass
of the order of $M\sim10^{-3}$eV and a self coupling constant $\lambda\sim10^{8}$.
They form stars with the mass of a mountain and so small as a football ball,
therefore they will behave as CDM and have the same problems as CDM. In this work we
want to avoid these problems, here
we will suppose that the mass $M$ and the self-interaction's parameter $\lambda$ are
very small, for example $M$ might be of the order of eV, like the neutrinos and
$\lambda\sim10^{-6}$, such that the critical mass of collapse is like the halo of a
galaxy. Thus, each BEC star made of this scalar field will form a galaxy. That
means, the main difference of this hypothesis with the axions is just this, namely,
the CDM galaxies form by collapse of dust made of a heavy or a heavy clump of
particles, while the scalar field galaxies form by condensation, the scalar field
freezes and form the halo of a galaxy. Thus, the important difference between these
two paradigms is that the galaxies in CDM have a density profile close to the center
of the galaxy like $\rho\sim1/r$, where $r$ is the distance measured from the
center. The CDM galaxies have a cusp central density profile. It is well know that
the BEC is completely regular at the center, in this context they generate core
density profiles. 

On the other hand, with this coupling constant and mass, scalar fields do not have
any problem with CMB observations. This, together with the fact that the scalar
field behaves like dust, i.e., like CDM at cosmological scales, guaranties that the
mass power spectrum and the angular power spectrum are the same as in the CDM
model.

Nevertheless, there is a second fundamental difference between CDM and this model.
Galaxies are hierarchically formed in the traditional CDM model, small galaxies
merger with other ones and form bigger galaxies, till they get the size they have
today. While they are formed by condensation in the case of the scalar field. It
means that in the scalar field paradigm galaxies will form very early, at least much
earlier than in the case of the CDM model. Thus, if the scalar field with these
parameters is the DM in the universe, we have to see well formed galaxies at high
redshifts and they must be core in the center, while if some particle like the WIMPs
is the DM, we have to see that galaxies form from redshifts $z\sim 6$ and they must
have a cusp central density profile. Summarizing, the scalar field and the CDM
models behave in the same way at cosmological scales, but at galactic scales they
have some differences. Two of these differences are that scalar field halos are core
and CDM ones are cusp and that scalar field halos form much earlier than the CDM
ones.

%We could obtain the same ranges for those parameter's values from relic density
%estimations.
%$\lambda \gtrsim \sqrt{\frac{m_W}{M_{Pl}}} \sim 10^{-8}$.

\section{Particle physics contrains}

Since the rise of the SM most of the proposals to extend the model have
been based on generalizations and/or higher symmetries which could 
include new fields in a constrained scheme containing the basic features of the SM.  
The recent negative results in the search for several constrained SUSY models and
scenarios in the LHC have changed the idea that
the best theoretically motivated models are the most appealing and have raised
serious doubts if, for example SUSY, cure more problems than those 
which create. The alternative approach is to propose an entirely phenomenological model solving
the DM problem and consistent with all the experimental
data available (astrophysical, cosmological, from colliders, etc.) and verify if is
it falsifiable.   The present idea it is not new, it has been proposed several times under different
names (darkon, phion, little higgs etc.) and in the current situation of
very new data measurements from LHC and Plank, the review of this simplest model is
compulsory.

\subsection{Vacuum and symmetry breaking pattern}

%%%%%%%%%%%%%%%%%%%%%%%%%%%%
%Question: Why most of the previous works include directly a 2-Doublet Structure for
%the scalars ? Because of SUSY ?
%%%%%%%%%%%%%%%%%%%%%%%%%%%%

The first condition for this model is that it must not have a visible influence on
the way the Standard Model Higgs breaks the EW 
symmetry. The relations that have to be held in order to have the new potential bounded from
bellow to ensure the existence of a 
vacuum can be seen elsewhere. The conditions to preserve the role of the SM Higgs in the EW symmetry break pattern
are 
$<S>=0$, $0<-m_0^2<v_{EW}^2\sqrt{\lambda_h\lambda_S}$ . If we use the conventional
shift to the
vacuum value for $h$ to represent the physical Higgs having mass $m_h^2=\lambda_h
v_{EW}^2$ 
the potential dependent of the new scalar can be written as

\begin{equation}
V=\frac{1}{2}(m_0^2+\lambda v_{EW}^2) S^2 +\frac{\lambda_{S}}{4}S^4+\lambda
v_{EW}S^2h+\frac{\lambda}{2}S^2h^2
\label{eq:potential}
\end{equation}

with the $S$ mass being $m_S^2+m_0^2+\lambda v_{EW}^2$. 
The prejudice some years ago was that the
range for this mass was from a few to a few hundred GeV in order to have cold dark
matter, this is not a 
consensus anymore.
There is another prejudice from the latter equation, in the cases for very low
masses for the new
scalar there is a new fine tuning to be explained. Being not the first or the most
relevant fine tuning
in field theory the model can allow it.

%\subsection{Limits from Galactic Halo formation}

\subsection{LHC Measurements}

There is a wide program in both LHC experiments, ATLAS and CMS, to search for Dark
Matter even if it can 
not be seen inside the detector. These searches rely in the capabilities to measure
the Transverse Missing 
Energy and reconstruct the events that have the DM at the end of the cascade. The
masses we are estimating
for the model are far from the best sensitivity of both experiments. Nevertheless
the scalar proposed
couples with the Higgs Boson and should be part of its invisible width. The
measurements of the 
Higgs boson couplings at the LHC allow to constrain the contribution of an invisible
decay to the 
total width. 
 
An extensive fit to the recent data ha been performed by Espinosa {\it et al.}
\cite{espinosa}
where they obtained that $Br_{inv}<0.37 (0.40)$ for $m_h=125 (126)$ GeV at 95$\%$
C.L.
The best fit gives a value of $Br_{inv} = 0.05 $ with 95$\%$ C.L.

The $h\rightarrow SS$ decay width is given by
\begin{equation} 
\Gamma(h\rightarrow SS) = \frac{1}{8\pi}\frac{\lambda^2
v_{EW}^2}{m_h}\sqrt{1-(\frac{2m_S}{m_h})^2}.
\end{equation} 

Using either the best fit or the limit  for the branching ratio of the Higgs'
Invisible width and the
expression of the Higgs' width to scalars we can find the allowed region for
coupling and mass combinations.

The limit on the Higgs' Invisible width can be interpreted as

\begin{equation} 
 \lambda^2 \sqrt{1-\frac{4m_s^2}{(125\:GeV)^2}} < 76.83 \:eV
\end{equation} 

and it is not a closed region(see Fig. 1).

\begin{figure}[h]%[tbp]
%\centering
\epsfig{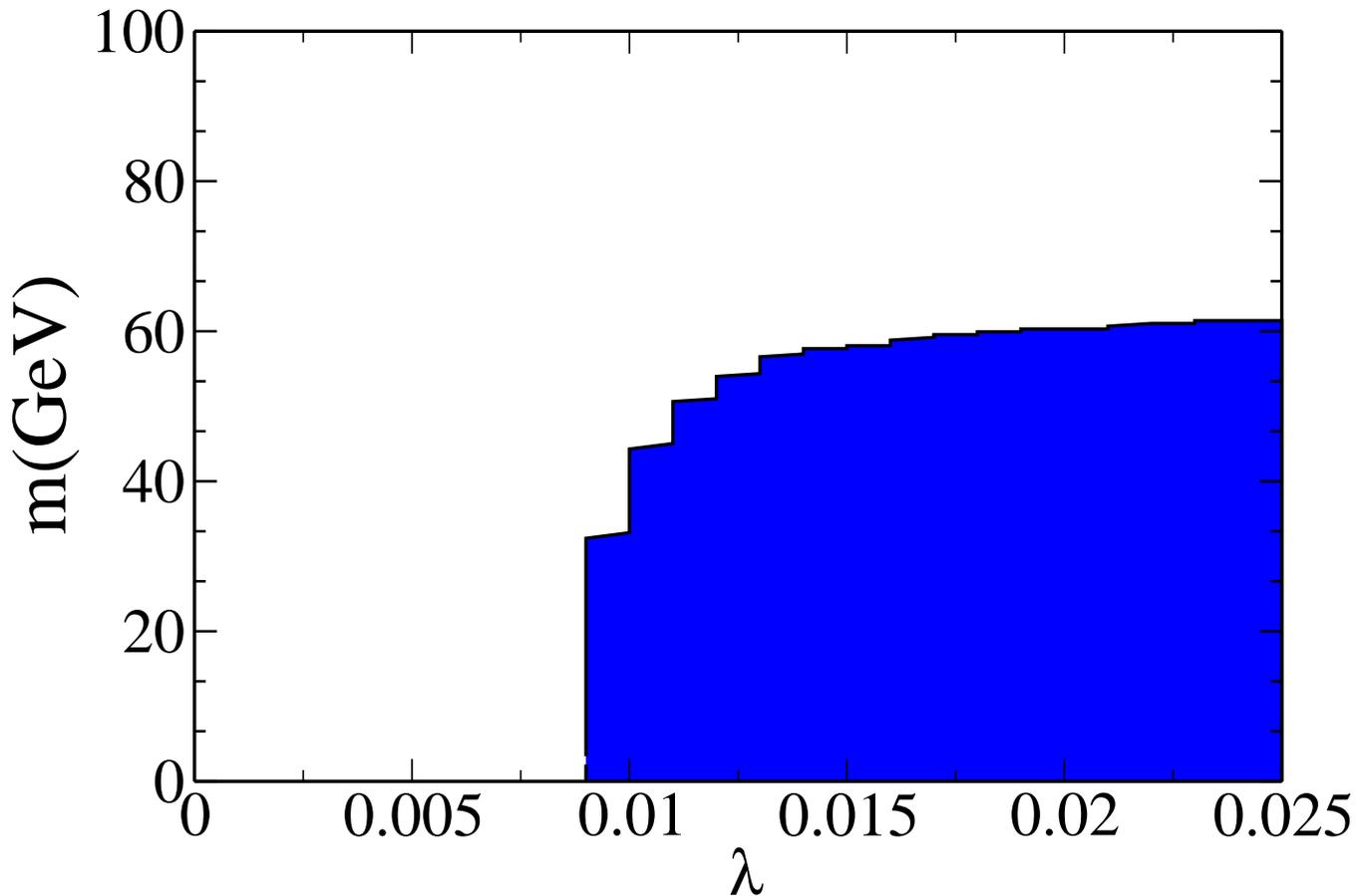}
\caption{Exclusion region (in blue) in the $\lambda$, mass plane from the Higgs
invisible width.}
\label{}
\end{figure}

We can use the mentioned limit from relic density on the coupling $\lambda \ge
10^{-8}$ to obtain
the limit on $m_S > 62.5$ GeV.

\subsection{Dark Matter Direct Searches}

The nucleon-scalar elastic scattering cross section can be also obtained for a
parameters space
region and compare it with the current limits from WIMP Searches.

From \cite{burgess} we can verify that the expression 

\begin{equation} 
\sigma_{el} (nucleon) \approx  \lambda^2  \left(\frac{100
GeV}{m_h}\right)^4\left(\frac{50 GeV}{m_S}\right)^2 (20 \times 10^-42 cm^2)
\end{equation} 

could be useful to find independent allowed combinations of $m_S$ and $\lambda$.

Recently some of the proposed experiments have quoted to plan to reach $10^-43$
cm$^2$ 
for this cross section. This cross section corresponds to 
model dependent parameter combinations
specific exclusions.  
 
%(I have to check the regions)

%\subsection{Possible relation with Dark Energy}

%The Higgs-Saw mechanism \cite{krauss} could be suited for the new scalar low mass
%and the big scale of the mixing with the SM. In this scheme, the coupling  could 
%naturally be negative representing also a Dark Energy candidate in that case.

%\section{Superposition of all available data}

%Some of the previously mentioned limits are consistent with each other but some
%other are completely out of the ranges. Most of the arguments are conservative but
%not all measurements have a high accuracy level. Can we make a combination ? 

\section*{Acknowledgments}
 
This work was partially supported by CONACyT M\'{e}xico
under grants CB-2009-01, no. 132400, CB-2011, no. 166212,  and I0101/131/07 C-
234/07 of the Instituto Avanzado de Cosmologia (IAC) collaboration
(http://www.iac.edu.mx/). 

%\bibliographystyle{<apsrev4-1>}

\end{document}